\documentclass[10 pt]{reu}
\usepackage{verbatim}
\usepackage{amsmath}
\usepackage{amssymb}
\begin{document}
\author{Mason A. Porter}
\address{Department of Physics and Center for the Physics of Information, California Institute of Technology, Pasadena, CA  91125, USA}
\email{mason@caltech.edu}
\title{Life on Both Sides of the Fence: Mentoring Versus Being Mentored}
\date{\today}
\maketitle




\section{Introduction} \label{intro}

The responsibility of scientists includes not only the development of new knowledge but also the education of future generations of scientists.  One of my own great pleasures has been the opportunity to mentor over twenty undergraduate students in various interdisciplinary projects in theoretical and computational science.  While my own research borders most closely with applied mathematics and theoretical physics, my goal is not to train mathematicians per se but to train {\it scientists}.  

My mentoring style is a direct product of my research, teaching, and social experiences during my undergraduacy at Caltech, from which I obtained a B.S. degree in applied mathematics in 1998.  
Caltech undergraduates are expected to drink from the firehose of knowledge, and I try to give the same opportunities to my students.  Of course, how much science I expect a research advisee to swallow depends greatly on the student.  Nevertheless, in order to help a student really appreciate science, it is extremely important to impart not only knowledge that is directly germane to their project and how to attack it (and other research problems), but also to convey just how much wonderful stuff there is for us to study (in any field!) and some of the places where the student's particular research problem fits into the big picture.

In the rest of this article, I expand on these ideas a bit.  I start by giving specific details on how undergraduate research is organized at Caltech.  I then briefly discuss my own undergraduate research experiences, commenting a bit on what I did and did not like.  I subsequently discuss my experiences as a mentor at Cornell (in the Mathematical and Theoretical Biological Institute), Georgia Tech, and Caltech and summarize with a few pithy pieces of advice.

\section{Undergraduate Research at Caltech}

Caltech's primary research program for undergraduate students is the Summer Undergraduate Research Fellowship (SURF) program, which has been around since 1979 and presently offers research opportunities---both on campus and at the Jet Propulsion Laboratory (JPL), which is run by Caltech---for about 250 Caltech students and 150 students from other institutions each year.  

The SURF program, a campus-wide research initiative, permeates Caltech's culture.  It encompasses all academic fields and is a fundamental part of Caltech's undergraduate education, as roughly three quarters of all Caltech undergraduates students participate in SURF at least once before they leave.   Moreover, one of Caltech's primary selling points to prospective undergraduates is the prominence of Nobel Laureates, National Academy members, and---more generally---the top scientists in just about every field in their classroom and research education from the very beginning of their undergraduate careers.  

As described at {\it http://www.surf.caltech.edu/}, the SURF program is modeled on the grant-seeking process:
\begin{itemize}
\item{Students collaborate with potential mentors to define and develop a project.}

\item{Applicants write research proposals for their projects.  (A prospective SURF student typically writes the entire proposal; his/her mentor may provide suggestions during the revision process.)}

\item{A faculty committee reviews the proposals and recommends awards.}

\item{Students carry out the work over a 10-week period in the summer (from the middle of June to late August), earning a salary of \$500 per week.}

\item{In order to ensure that students keep the big picture in mind, they are required during the summer to submit two progress reports that detail unexpected challenges, how their goals have changed from the description in their project proposal, and other similar items.} 

\item{At the conclusion of the program, students submit a technical paper and give an oral presentation at SURF Seminar Day, a symposium modeled on a professional technical meeting.  There is a seminar day predominantly for non-Caltech students in the middle of August and one for Caltech students in the middle of October.  A draft of the final report is due right before Caltech's fall quarter starts at the end of September, but the final version approved by the mentor is not due until about November 1st.}
\end{itemize}

Every year starting in December, the entire Caltech faculty is solicited to advertise SURF projects that they are offering for the following summer.  As responses are received, research opportunities are compiled on a website, which is organized according to student majors (or, more precisely, according to Caltech \textquotedblleft Divisions" such as \textquotedblleft Physics, Math, and Astronomy").\footnote{See {\it http://www.surf.caltech.edu/applicants/opportunity/index.html}.}  Each opportunity includes a terse description of the problem(s) available and other germane information such as allowed majors; coursework requirements; whether the work is theoretical, computational, or experimental in nature; whether non-Caltech students will be considered, etc.  Students then contact the appropriate professor (or perhaps even ones who have not advertised projects, which is a lesser-known but often successful strategy) and arrange to meet with them.  Typically, potential mentors have informal interviews with several students for a given project before they decide which one(s) they want to advise.  Mentors who are postdoctoral scholars or graduate students must also have a faculty sponsor listed on the SURF opportunity.


To enrich the students' research experience, Caltech's SURF program includes a variety of activities aside from individual research projects---including weekly (accessible!) seminars by Caltech faculty and JPL technical staff; a participatory discussion series on developing a research career, graduate school admissions, and other topics of interest to future researchers; and social and cultural activities sponsored by the SURF Student Advisory Council.  These activities facilitate interactions between students coming from other universities and those who attend Caltech, which tends to have a fairly insular student body.

\section{My Undergraduate Research Experiences}

As a Caltech undergraduate, I undertook three research projects.  Here I'll briefly discuss the two that were through the SURF program.
Although many projects are advertised, both of mine arose from unsolicited e-mails I sent to professors.\footnote{I have since learned that many faculty never advertise projects and only advise students who take the initiative to contact them.}


During my freshman year, I attempted unsuccessfully to obtain a SURF project on cellular automata with a mathematics postdoc.  However, he was willing to supervise an independent reading course in tensor analysis and geometry using a book by Jerry Marsden, who he told me would be joining Caltech as a faculty member in the fall.  
Armed with two of Marsden's books---I used his analysis book in a course I took at UCLA that summer---and slightly star-struck, I arranged to meet with him once the school year started in order to get my books autographed.  Later, when it came time to apply for summer 1996 SURF projects, I looked at the advertised projects and wasn't interested in any of them.  I already knew that I wanted to have a research career in dynamical systems, so I contacted Marsden 
and ended up working with him on a project entailing the writing of an expository article on the Hopf fibration and its application to problems in mechanics.
 That summer, I learned not only a lot of geometric mechanics but also a great deal concerning how to write mathematical articles and how to use \LaTeX.

I also learned a few other lessons from this first research experience.  For example, I tried to get the expository article that resulted from my project published in various venues, although I was never successful.  It is worth noting that I did this entirely on my own when what I should have done was convert my manuscript to a coauthored article and ask Marsden to help me find an appropriate place to submit the paper.  (I did ask his opinion about where to send the paper, but I pursued the submission process on my own.)
While it is probably very rare for undergraduates to take such initiative, I have made it a point to discuss the publication and dissemination of work (in some context) with all my research students, if only to make sure they don't make the same mistake I did.  I have published papers with several of them, but I wanted them to know that that was something we would do together.


My summer 1997 SURF project, which I also obtained by contacting a professor
who had not advertised a project, taught me a lot about my own interests and what {\it not} to do as a mentor.  
During our first meeting, my advisor-to-be offered me my choice of a computational or theorem-proof project.
I felt it would be useful to try my hand at research in pure mathematics and when the professor told me that numerous prior students had tried the same project without success, I knew I had to try it.  Unfortunately, that project was a complete disaster for me as well.  I did learn a lot and had the chance to read some seminal physics papers, but my advisor was absent from campus virtually the entire summer and I was basically stuck the entire time with nobody to ask for help.\footnote{By contrast, Marsden was often reachable when he was out of town and was also very explicit about when he would be gone, so while he might be away from e-mail for a few days, during my second SURF project, I wouldn't get a response for several weeks at a time (for a ten week project).}  I didn't yet know how to find journal articles on my own or even which journal articles I should get, what techniques I should try to learn, or (honestly) much of anything.  
I was interested in the physical phenomenon (diffusion limited aggregation) about which I was supposed to be proving a theorem, so one major thing I learned was that pure mathematics was not for me.  Naturally, I need to know a lot of pure mathematics to do research in applied mathematics, but in my own research I have come to insist on close ties to physics or other application areas.  Hence, my summer 1997 project played an important role in my career by making me appreciate a few of the things I did {\it not} want.  It also complemented the previous summer by teaching me a few things that a mentor should not do: {\it While sometimes one's students who show more results and ask more questions may deserve more attention (which is perhaps counterintuitive), that does not mean you should let the others sink or swim.  Instead, teach them to ask questions and think creatively even if they have trouble learning the science.}  I also decided that undergraduates should be taught how to search the research literature as part of their research experience.


Although I did not publish any papers from my undergraduate research, my various experiences played strong roles in shaping my mentoring style.  I also learned a lot of technical material, how to use \LaTeX, how to write scientific reports, and what to do and not to do as an advisor.


\section{The Transition from Mentee to Mentor}

\subsection{The Mathematical and Theoretical Biology Institute (MTBI)}


My first significant mentoring experience came during the summers of 2000 to 2002 as part of Carlos Castillo-Chavez's Mathematical and Theoretical Biological Institute (MTBI).\footnote{See {\it http://mtbi.asu.edu/}.}  My role included both assisting students with homework problems and helping out as an advisor and critic for research projects.  

Because of the benefits I had already gained by attending conferences, I strongly encouraged the MTBI students to attend research conferences.  In 2002, the Society for Industrial and Applied Mathematics's annual summer meeting was held in Philadelphia.  This allowed us to arrange a road trip so that the current MTBI students could attend the meeting, where Stephen Wirkus and I co-organized two sessions on mathematical biology whose speakers were all MTBI alums \cite{generation}.  This \textquotedblleft minisymposium" allowed the students to see what people who used to be in their shoes had accomplished, talk to them about both academic and social issues, and experience the usual benefits of attending conferences.

At MTBI, I was known for my tendency to ask tough questions during final presentations and return manuscript drafts containing numerous suggestions written lovingly in red.  My experience at Caltech taught me that academic rigor should be stressed from the start.  Moreover, there are times when one should be blunt with students in order to improve certain aspects of their work.  For example, I think it's extremely important to ask tough questions in a relatively friendly environment to prepare students for future situations, such as doctoral thesis defenses and conference talks, where the audience might not be a priori favorably disposed towards their research.  On occasion, students would try to finagle an answer to my questions, which would invariably lead to even tougher follow-up questions in order to teach them that that is simply not permissible.  I find that many students don't appreciate that it is acceptable, and in fact preferable, for them to give an honest answer of 'I don't know.' and (ideally) to ask the questioner to discuss the matter further offline.

My first chance to be the main advisor on an MTBI research project was an important learning experience for me.  I felt my regular meetings with the students were reasonably productive, but mentors and students need to have personalities that are at least somewhat compatible.  In this particular case, I had a personality conflict with a student who I found to be particularly abrasive and whose academic background was extremely poor.  My initial expectations for her were too high, and she did not appreciate my attempts to push her.  Thus, while my belief that one should have high expectations for every student was untempered, I realized that what actually constitutes this varies greatly from student to student.  Essentially, my goal became to make sure that students left my charge in a more advanced state than when they entered it.  For some students, this means a peer-reviewed publication, but for others it may be as simple as understanding (and having duplicated on their own) some calculations, proofs, or numerical computations from a book or article.  Usually, things fall somewhere in between these extremes.  I also learned from that project that it is essential that I show my frustration only when I think a kick in the butt will help the student.  I still experience considerable frustration when advising many of my students, but I am purposely better at showing it only when I have judged it's what the student needs to see.  (Whether that judgement is correct in a given instance is obviously something that can be debated.)  The third thing I learned from this experience concerned what projects I wanted to advise: For me to be a good mentor, it is imperative not only that the students be interested in the project but also that I am as well.  I didn't particularly appreciate this limitation before that experience.

\subsection{Georgia Tech's VIGRE Program}


I began advising projects at Georgia Tech in summer 2003 through the math department's REU, which was funded through the department's National Science Foundation VIGRE grant.\footnote{A slightly outdated version of the format for these projects is available at {\it http://www.math.gatech.edu/$\sim$lacey/ump/reu/reu.htm\#vigre}.  The projects that have been undertaken are listed at {\it http://www.math.gatech.edu/$\sim$lacey/ump/reu/gtreulist.htm}.}  

When a student first contacts me about research, we meet in person so that I can find out his/her background and interests.  In my experience, many students are ready to undertake research even without an extensive coursework background, and I like to be able to accommodate that.  (In fact, some of my best students initially approached me during their freshman years.)  Alternatively, one can adjust the scope of projects to accommodate students who haven't had as many courses.  Things might progress slower, but the first term of advising a project typically entails (with all but the best students) more time learning background material than producing original results anyway.  During our first conversation, I also encourage my prospective advisees to contact students who have worked with me in the past.  Most students seem to be ready to sign up on the spot, but I want them to get into the habit of seeking advice from those who have gone before them.\footnote{This is the same advice that is invariably passed on to new doctoral students when it comes time for them to find Ph.D. advisors.  Students are {\it extremely} honest when it comes to talking about current and former advisors.}  Student projects that lead to publications invariably last longer than one term/summer.  Indeed, the publications I have obtained from student mentoring have all been co-authored with students who worked with me for two--four semesters rather than just one.


My ideas for projects come from several sources.  One of them, on modeling bipolar disorder, was a continuation of an MTBI project I had advised in summer 2002.  Another, involving the construction of a graphical user interface to simulate billiard systems, was motivated by a question that arose during a seminar at the Mathematical Sciences Research Institute in which an audience member wondered how everybody was making the graphics for the talks in a quantum chaos workshop.  I have also obtained many others through traditional means such as the research literature.  Many of these have been in network theory, a subject that is particularly suited for undergraduate research.\footnote{There are very few theorems that apply to real-world networks, the importance of many of the problems is easy to explain, and it does not take very much coursework to be ready to conduct research in this subject.}  

On several occasions, I co-advised student research projects (with, for example, two students working on similar projects) with Georgia Tech faculty to facilitate group meetings, establish collaborations, and delve into new research areas.  Starting in summer 2003, for example, I co-advised a pair of students (with assistant professor Peter Mucha) who were studying two different networks (Congressional committee assignments and the NCAA Division-IA football schedule) but often needed to learn similar concepts and compute similar quantities.  I found that students typically advance much faster when they are working on similar projects and can bounce ideas off each other.  Assigning two students to the same project can be beneficial as well, but I think it's much better if their projects overlap rather than duplicate each other.  Additionally, the presence of two advisors with complementary skills---Peter specializes in scientific computation and has done most of his research on problems in fluid mechanics---is wonderful for both the students and the mentors (and is ideal for collaborations in general).  It is also worth noting that while I had learned about network theory through coursework and reading research papers, I had never actually done any research in that area and was very keen to do so.  These research projects were thus my first forays into network theory (in which I have remained active).\footnote{More recently, my network-theory collaborators and I have instituted a research-group wiki ({\it http://www.unc.edu/$\sim$mucha/netwiki/}), where we hope to post papers we would like our students to read, concepts we think they should understand, links to introductory material on software like Matlab and \LaTeX, and other useful material.  It would be wonderful if wikis devoted to undergraduate research---with separate subpages for each REU---are developed.}  My subsequent research groups have included students from multiple majors and faculty from other departments, allowing my students to be introduced not only to research but also to interdisciplinary collaboration.  (For applied mathematics projects, it is ideal to involve faculty from multiple departments.)  For example, one group I initiated included a biology professor, a math professor, an electrical engineering major, a math major, and me.

Motivated by Caltech's framework, I insisted that my students write final reports that we would then polish together by working through several drafts.  (I also insisted that they use \LaTeX to write these reports and provided them with a tutorial for first-time users that I had written while at MTBI. \cite{lala})  This not only improved my students' communication skills but also allowed us to better appreciate the gaps that still needed to be filled in when it came to possible publication.  I also arranged for my students to give short talks in group meetings of Georgia Tech's Center for Nonlinear Science, with which I was affiliated.  Moreoover, I encourage my students to present posters and talks at conferences and occasionally took them with me to meetings, ranging from Ohio State's Young Mathematicians Conference\footnote{See {\it http://www.math.ohio-state.edu/conferences/ymc/}.} to research conferences such as Dynamics Days.\footnote{See {\it http://www.bu.edu/provost/ddays\_07/} for the homepage for Dynamics Days 2007.}

Perhaps a bit more unusual, I organized a public lecture (as well as a more technical seminar) at Georgia Tech by Cornell applied mathematician Steve Strogatz, which gave my students (some of whom had used his books and/or articles to help them with their research) a chance to meet him.  
While not many mathematicians give public lectures, one can still ask colloquium speakers to also give more general talks for undergraduates (or design the colloquium series itself to include such talks) and arrange for them to have lunch or otherwise meet informally with undergraduates.  I remember attending several lunches with colloquium speakers as a graduate student, and the ensuing informal discussions give students opportunities to be heard that are not otherwise available.  

\subsection{On the Other Side at Caltech}



I have continued my student mentoring as a Caltech postdoc.  The 2006 version of my research-opportunity announcement read as follows:

\begin{verbatim}
Projects in Nonlinear Dynamics and Complex Systems

Michael Cross, Professor
Department of Physics

Mason Porter 
Postdoctoral Scholar
Department of Physics and Center for the Physics of Information 

mason@caltech.edu
130 Sloan Annex

Majors: Any major is good, but students in applied math, physics, 
math are likely to be especially interested in these projects.

Prerequisites: It depends on the specific project. Some require more 
than others. I will work with the student's background to design something 
appropriate. This should not be considered any sort of obstacle.

Type: Theoretical and/or Computational

Note: This is an on-campus SURF. Caltech students only.

Nonlinearity and complexity abound throughout science, nature, and 
technology, as their understanding helps to provide explanations of 
myriad phenomena---including synchronization of flashing fireflies and 
lasers, chaotic motion in double pendula, the formation of patterns in 
chemical reactions, species co-existence in plankton populations, correlations 
between political ideology and congressional committee structure, and 
chaotic dynamics in both classical and quantum systems. In this project, 
the student(s) who work with me will work on some mathematical modeling 
of some phenomenon (to be discussed in private communication) using 
computational and/or analytical techniques. To get an idea of the types of 
things I like to study, please see www.its.caltech.edu/~mason/research or 
drop me a line. Possible projects include ones involving Bose-Einstein 
condensates, complex networks (such as congressional networks), quantum 
chaos, billiard systems, pattern formation, synchronization, and others.
\end{verbatim}

Several of these items require some explanation.  First, a member of the Caltech faculty (in this case, my postdoctoral advisor Michael Cross) must be listed as the official mentor.  
Caltech asks mentors to list \textquotedblleft allowed" majors, although I instead indicated who was more likely to enjoy my projects.  I purposely included several project ideas with the intention that the particular projects students would choose would be a function of not only my interests but also theirs.\footnote{In one case in 2005, one prospective advisee 
was sufficiently advanced
that I asked him to come up with his own project, which proved to be very successful.}

I require my students to write four or five drafts of their SURF proposal before I am willing to sign off on it, and I make this requirement clear when we first meet to discuss possible projects.
I show them sample proposals written by former students to use as models.  In some cases, I have also shown my old proposals to some of my students to accentuate the fact that I was in their position not so long ago.  Because they have seen my recent professional writing rather than my rougher undergraduate work, it is very useful as positive feedback after all the time they spend on their proposals to show them that their proposals are better than mine were. 

One change I have made in my advising since returning to Caltech has been the establishment of group meetings along the lines of what is perhaps more familiar in physics departments than in math departments.  Hence, in addition to meeting individually with each student twice a week (for roughly 30 minutes each time) and communicating over e-mail and instant messaging, my students and I all meet for about an hour every week.\footnote{This is similar to what our network theory group did at Georgia Tech, except the scope of the projects under investigation is larger.  All my students are involved rather than just ones in the topics that are most intimately related.  This allows them to learn more about nonlinear dynamics and complex systems beyond what is directly involved in their project and to field questions from a broader audience and give/receive more feedback from their peers.}  In a typical meeting, two of my students will present material in front of the whole group, the others are expected to ask questions about concepts they don't understand, and I will occasionally speak up when I want to highlight a particular point (which may be related to science, exposition, or both).  In some meetings, students practice more formal oral presentations or read and discuss each others' written reports.  
I specifically instruct my students to ask tough questions and make critical comments, and
I expect my them to be similarly critical when reading their peers' papers, as it is imperative that they understand that they are not doing their friends a favor---in fact, they are doing them a disservice---if they go easy on them.  While my purpose with these exercises is to improve their oral and written presentation skills, I think the students also might learn a bit about peer review as a byproduct.  Once my students have given comments, I add a few of my own, and I accentuate the previously-raised points with which I particularly agree.  (My students invariably bring up nearly all of the most essential points I intend to raise, and the fact that multiple people are noticing the same things reinforces the ensuing suggestions.)

The final SURF reports are supposed to be written in the format of an appropriate scientific journal.  Caltech's SURF office recommends {\it Nature} as a default but leaves the final decision to the research mentors.  I discuss publishing issues with all my students (even the weaker ones) and select a journal style that is appropriate for their particular project and situation.  My past journal models have included not only broadly-oriented venues such as {\it Nature} and the {\it Proceedings of the National Academy of Sciences} but also archival journals like {\it Physical Review E} and {\it Chaos}.  As with the initial proposal, I go through several drafts of the final report with each student.


\section{Conclusions}\label{conclusion}

As I have discussed at length, my undergraduate research experiences at Caltech and my interdisciplinary training have fundamentally shaped my mentoring style.  In my opinion, students obtain an optimal research experience when expectations are high---they should drink from the proverbial firehose of knowledge---but not too high.\footnote{This is a variant of a quote by Albert Einstein: \textquotedblleft Make everything as simple as possible, but not simpler."}  It is extremely valuable for students and faculty from multiple backgrounds to interact regularly, and weekly group meetings provide an excellent supplement to one-on-one meetings by reinforcing this and other boons.  From a selfish perspective, I have found advising student projects to be an ideal means to enter new research areas.  Although it can be very frustrating at times, my close involvement with talented students is perhaps my favorite academic pursuit.

\section*{Acknowledgements}

I gratefully acknowledge Stephen Wirkus and Ed Mosteig for reading an early version of this manuscript and giving me numerous helpful suggestions.


\begin{thebibliography}{1}

\bibitem{lala}
{\sc M.~A. Porter}, {\em A {H}itchhiker's {G}uide to {L}a{T}e{X}: Or how I
  learned to stop worrying and love writing my dissertation}.
\newblock {\it http://www.math.gatech.edu/$\sim$mason/papers/draft/lala.pdf},
  2002.

\bibitem{generation}
{\sc S.~Wirkus and M.~A. Porter}, {\em {SIAM} hears from next-generation
  mathematical biologists at {P}hiladelphia meeting}, SIAM News, 35 (2002).

\end{thebibliography}

\end{document}